\documentclass{appolb}
\usepackage{graphicx}
\begin{document}
\title{Superconductivity and Instabilities in \\ the $t-t'$ Hubbard Model
\thanks{Presented at the Strongly Correlated Electron Systems 
Conference, Krak\'ow 2002}%
}


\author{F.~Wegner$^{(1)}$ and V.~Hankevych$^{(1),(2)}$
\address{$^{(1)}$ Institut f\" ur Theoretische Physik, Universit\" at 
Heidelberg, Philosophenweg 19, D-69120 Heidelberg, Germany\\
$^{(2)}$ Department of Physics, Ternopil State Technical University, 
56 Rus'ka St., UA-46001 Ternopil, Ukraine}
}
\maketitle


\begin{abstract}
We present a stability analysis of the 2D $t-t'$ Hubbard model on a square lattice for $t'=-t/6$. We find possible phases of the model ($d$-wave Pomeranchuk and superconducting states, band splitting, singlet and triplet flux phases), and study the interplay of them. 
\end{abstract}


  
\section{Introduction}
In recent years the two-dimensional (2D) Hubbard model has been used as the simplest model to describe  
the electron correlations in the copper-oxide planes of high-temperature superconductors since experimental data suggest that superconductivity in cuprates basically originates from the CuO$_2$ 
layers. 
Apart from the antiferromagnetism and $d_{x^2-y^2}$-wave superconductivity, 
 a few other instabilities related to symmetry-broken 
states~\cite{am,ko,hm,clmn,gkw,hgw} 
and occurring together with them in the 2D $t-t'$ Hubbard model with 
next-nearest-neighbor hopping $t'$ have been reported. They are the flux phase\cite{am,ko} or $d$-wave density order~\cite{clmn}, the triplet flux phase~\cite{hgw}, the $d$-wave Pomeranchuk instability~\cite{hm} and band 
splitting~\cite{gkw}. Ferromagnetism and 
$p$-wave triplet superconductivity have been observed also by the authors of Refs.~\cite{hsg,foh,ikk,hs} and Ref.~\cite{hs} respectively at certain region of electron concentration around the Van Hove filling (where the Fermi surface passes through the saddle points of the single particle dispersion) for large negative values $t'$.

However, the competition and interplay of these phases remain an open problem. In this paper we investigate superconducting and other possible instabilities of the 2D $t-t'$ Hubbard model at small negative value of $t'$. 
We consider also the leading instabilities depending on the ratio $U/t$ (in the papers cited above it was fixed). 

We start from the Hamiltonian of the $t-t'$ Hubbard model 
\begin{eqnarray}
H=\sum_{{\bf k}\sigma}\varepsilon_{\bf k}c^\dagger_{{\bf k}\sigma}c_{{\bf k}\sigma}+
{U\over N}\sum_{{\bf k_1 k_1' \atop k_2 k_2'}}c^\dagger_{{\bf k_1}\uparrow}c_{{\bf k_1'}\uparrow}c^\dagger_{{\bf k_2}\downarrow}c_{{\bf k_2'}\downarrow}\delta_{{\bf k_1}+{\bf k_2},{\bf k_1'}+{\bf k_2'}},
\end{eqnarray}
where $\varepsilon_{\bf k}$ is the Bloch electron energy with the momentum ${\bf k}$, $c^\dagger_{{\bf k}\sigma} (c_{{\bf k}\sigma})$ is the creation (annihilation) operator for the electrons with spin projection $\sigma \in \{\uparrow,\downarrow\}$, $U$ is the local Coulomb repulsion of two electrons of opposite spins, $N$ is the number of lattice points, lattice spacing equals unity.
For a square lattice the single particle dispersion has the form
\begin{eqnarray}
\varepsilon_{\bf k}=-2t(\cos k_x+\cos k_y)-4t'\cos k_x\cos k_y. \label{ek}
\end{eqnarray}

By means of the flow equation method~\cite{we} the Hamiltonian is transformed into one of molecular-field type. This Hamiltonian is calculated in second order
in the coupling $U$~\cite{gkw}.
Adopting the notations of Ref.~\cite{gkw}, free energy can be expressed by the
order parameters $\Delta_{\bf k}$ in the form:
\begin{eqnarray}
\beta F={1\over N}\sum_{\bf k q} \beta U\left(1+{U\over t}
V_{{\bf k}, {\bf q}}\right)\Delta_{\bf k}^{\ast}\Delta_{\bf q}+
\sum_{\bf k}f_{\bf k}\Delta_{\bf k}^{\ast}\Delta_{\bf k}, \label{fe}
\end{eqnarray}
where the first term is the energy contribution and the second term is the entropy contribution, $\beta =1/(k_BT)$, $T$ is the temperature, $t$ is the hopping integral of electrons between nearest neighbors of the lattice, $V_{{\bf k}, {\bf q}}$ is effective second-order interaction, and 
$f_{\bf k}$ is an entropy coefficient. All quantities of Eq.~(\ref{fe}) are defined in Ref.~\cite{gkw}. 

We start from the symmetric state and investigate whether this state is stable against fluctuations of the order parameters $\Delta$. As soon as a non-zero $\Delta$ yields a lower free energy in comparison with the symmetric state 
$\Delta\equiv 0$, then the symmetric state is unstable and the system will approach a symmetry broken state. This indicates a phase transition.

\section{Results and discussion}

We perform numerical calculation on a square lattice with $24\times 24$ 
points in the Brillouin zone for the various representations under the 
point group $C_{4\nu}$. 
Initially, such numerical calculations have been performed in 
Refs.~\cite{gkw,gr} for the 2D Hubbard model, but they were sensitive to the lattice size at low temperatures. Here we use an improved scheme (for details see Ref.~\cite{hw}).

\begin{figure}[!h]
\includegraphics{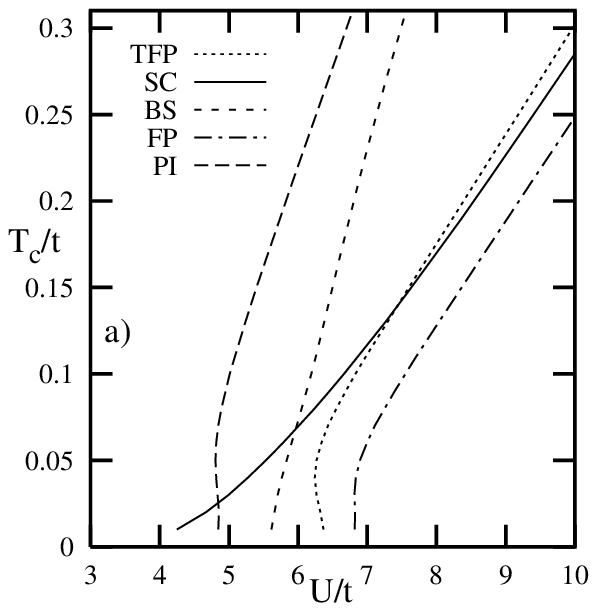}
\hfill
\includegraphics{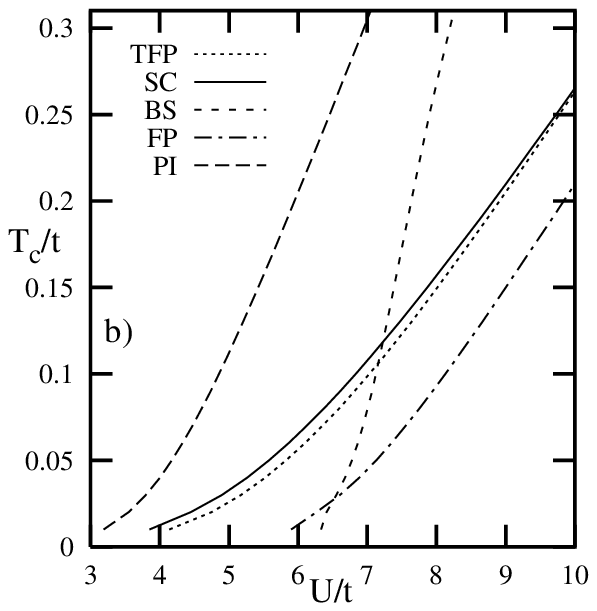}
\caption{Temperature phase diagram of the model at $t'=-t/6$ for $n=0.95$ (a) 
and $n=0.86$ (b). SC stands for superconductivity, BS for band splitting, PI for Pomeranchuk instability,  FP for flux phase, and TFP for triplet flux phase.}
\label{fig1}
\end{figure}

Apart from antiferromagnetism at small $t'$ and half-filling, one of the leading instabilities at small doping is a Pomeranchuk instability with $d_{x^2-y^2}$-wave symmetry in the singlet channel (see Fig.~\ref{fig1}). 
The corresponding eigenvector signals a deformation of the Fermi surface which breaks the point group symmetry of the square lattice from tetragonal to orthorhombic. At high temperatures the system has a tetragonal structure and an orthorhombic one at low temperatures. One can see from Fig.~\ref{fig1} that at the values $U\geq 6t$ the critical temperature of this transition decreases with increasing the hole doping $\delta \equiv 1-n$ ($n$ is the electron concentration). It means that the hole doping enhances the tendency towards an orthorhombic distortion of the Fermi surface (or lattice). The $d_{x^2-y^2}$-wave Pomeranchuk instability dominates at the Van Hove filling (Fig.~\ref{fig1}b).   

The next instability, which is developed in the region of electron concentration around half-filling and is one of the strongest in that region, is a particle-hole instability of singlet type with staggered $p$-wave symmetry. It yields~\cite{gkw} a splitting into two bands and may lead to an energy gap in the charge excitations spectrum. Other instabilities are the singlet and triplet flux phases. In contrast to the case of $t'=0$, where the singlet and triplet $T_c$ 
of the particle-hole instabilities with staggered symmetry of 
$d_{x^2-y^2}$-wave character are degenerate (that is the flux phase), they are different at $t'\neq 0$ and the triplet one is higher.

The superconducting $d_{x^2-y^2}$ instability is the strongest one at small doping and low temperatures. It is not destroyed at sufficiently large doping as well as large values of $|t'|$. At weak coupling $U<5t$ and close to half-filling the transition from a paramagnetic phase to superconducting one can occur at very low temperatures (Fig.~\ref{fig1}b). The peculiar feature of the superconducting phase should be noted. Away from the Van Hove filling (at the Van Hove filling the density of states has a singularity, Fig.~\ref{fig1}b corresponds to this situation) when temperature approaches zero 
the curves corresponding to superconducting phase are flat, whereas the curves corresponding to all other phases observed become steep. Therefore, at very low temperatures the transition from a paramagnetic phase to the superconducting one can occur at very small values of the corresponding effective interaction in contrast with the transitions to other possible phases which require some finite values of the effective interactions. One can see also that the critical temperatures of all phases increase with the increase of correlation strength $U/t$. Thus, electron correlations enhance the tendency towards the transition to the phases observed by us. 

In conclusion, we have presented a stability analysis of the 2D $t-t'$ Hubbard model on a square lattice. We have found possible phases of the model ($d$-wave Pomeranchuk and superconducting states, band splitting, singlet and triplet flux phases), and studied the interplay of them. 
One phase may suppress another phase. To which extend two order parameters can coexist with each other is a question, which has to be investigated in the future.


\end{document}